\documentclass[epj]{svjour}
\usepackage{graphics}
\usepackage[dvips]{color}%remove later
\begin{document}
\title{First measurements of the $^{16}$O(e,e$'$pn)$^{14}$N reaction}
\author{D.G. Middleton\inst{1}\thanks{\emph{Present address:}Physikalisches Institut, Universit\"{a}t T\"{u}bingen, Germany}\and J.R.M. Annand\inst{1} \and C. Barbieri\inst{2} \and P. Barneo\inst{3} \and P. Bartsch \inst{4} \and D. Bauman \inst{4} \and J. Bermuth \inst{4} \and D. Bosnar \inst{5} \and H.P. Blok\inst{3} \and R. B\"{o}hm\inst{4}  \and M. Ding \inst{4} \and M.O. Distler\inst{4} \and D. Elsner \inst{4} \and J. Friedrich\inst{4} \and C. Giusti \inst{6} \and  D.I. Glazier\inst{1} \and P. Grabmayr\inst{7} \and\newline S. Gr\"{o}zinger \inst{4} \and T. Hehl\inst{7} \and J. Heim\inst{7} \and W.H.A Hesselink\inst{3,8} \and E. Jans\inst{3} \and F. Klein \inst{4} \and M. Kohl\inst{9} \and L. Lapik\'{a}s\inst{3} \and\newline I.J.D. MacGregor\inst{1} \and I. Martin\inst{7} \and J.C. McGeorge\inst{1} \and H. Merkel\inst{4} \and P. Merle \inst{4} \and F. Moschini\inst{7} \and U. M\"{u}ller \inst{4} \and Th. Pospischil \inst{4} \and\newline G. Rosner\inst{1} \and H. Schmieden \inst{4} \and M. Seimetz \inst{4} \and A. S\"{u}le \inst{4} \and H.~de~Vries\inst{3} \and Th. Walcher\inst{4} \and D.P. Watts\inst{1,10} \and M. Weis \inst{4} \and\newline B. Zihlmann\inst{3,8}
%\thanks{}%
}                     % Do not remove
\institute{Department of Physics and Astronomy, University
of Glasgow, Scotland \and
 Gesellschaft f\"{u}r Schwerionenforschung, Planckstr. 1, 64291 Darmstadt, Germany
 \and NIKHEF, P.O. Box 41882, Amsterdam, The Netherlands \and Institut f\"{u}r Kernphysik, Universit\"{a}t Mainz, Germany  \and Department of Physics, University of Zagreb, Croatia \and Dipartimento di Fisica Nucleare e Teorica dell'Universit\`{a} degli Studi di Pavia and Istituto Nazionale di Fisica Nucleare, \newline Sezione di Pavia, I-27100 Pavia, Italy \and Physikalisches Institut, Universit\"{a}t T\"{u}bingen, Germany \and Vrije Universiteit, Amsterdam, The Netherlands \and Institut für Kernphysik, Technische Universität Darmstadt, D-64289 Darmstadt, Germany \and School of Physics, University of Edinburgh, Scotland}
\date{Received: date / Revised version: date}
\abstract{
This paper reports on the first measurement of the $^{16}$O(e,e$'$pn)$^{14}$N
reaction. Data were measured in kinematics centred on a super-parallel
geometry at energy and momentum transfers of 215 MeV and 316 MeV/c. The experimental resolution was sufficient to
distinguish groups of states in the residual nucleus but not good
enough to separate individual states. The data show a strong dependence
on missing momentum and this dependence appears to be different
for two groups of states in the residual nucleus.
Theoretical calculations of the reaction using the Pavia code
do not reproduce the shape or the magnitude of the data.
 % end of PACS codes
} %end of abstract
\maketitle
\section{Introduction}
\label{intro}
It is generally accepted that independent particle models, in which
nucleons move in a mean field, provide a good description of the structure of nuclei.
However, it has been shown that it is necessary to include two-nucleon
correlations beyond the mean field in order to describe some basic
nuclear properties such as binding energy \cite{IPM_srcs,NN_int_rev}.

Further evidence of the effects of correlated two-nucleon behaviour
comes from measurements of spectroscopic\newline strengths for the knockout
of protons from nuclei with A\,$>$\,4 \cite{low_shell_occupancy2_2,low_shell_occupancy,low_shell_occupancy3}.
The average value of the measured spectroscopic strength for a range
of nuclei was found to be $\approx$ 65\,\% of that predicted by independent
particle models. It was also found that the spectroscopic strength
of the `normally empty' orbitals above the Fermi edge was non zero.
The observed depletion of the Fermi sea is, to a large extent, believed
to be due to the influence of NN-correlations in the nucleus. These nucleon-nucleon correlations cause the promotion of nucleons to states above the Fermi level and generate large nucleon momenta within the nucleus.

Modern potentials describing the mutual interaction of nucleons contain many components, which are dependent on distance, spin and isospin of the nucleons, many of which are non-local. The potential at short distances is dominated by a strong scalar repulsive interaction, which prevents a nucleus from collapsing and which suppresses the uncorrelated nuclear wave function at small intranucleonic distances. These are the so-called short range correlations (SRC). A possible way to account for the SRC is the introduction of defect (i.e. suppression) functions for the radial wave functions, which are calculated for each partial wave by solving the Schr\"{o}dinger equation for two bound nucleons within a model space that contains the high-momentum degrees of freedom and accounting for the Pauli blocking effects of the remaining A-2 nucleons \cite{16O_eepN,16o_eepn_giusti}. It is predicted, that SRC contribute differently for different final states of the residual nucleus \cite{eeNN_theory_giusti}.

The second main component of NN-correlations is a tensor correlation
(TC) term, which depends on the spin and spatial orientation of the
two nucleons \cite{IPM_srcs,tensor_model}. This term favours energetically
the situation where the spins of neighbouring nucleons are aligned along their
relative vector separation. These correlations are mainly due to the
strong tensor components of the pion-exchange contribution to the
NN interaction and are very important in the wave function of a pn
pair, but much less so for pp pairs \cite{16O_eepN}.

Electromagnetically induced two-nucleon knockout is a powerful tool to investigate the role of correlated NN motion in the nucleus. A real or virtual photon, absorbed by one of the nucleons in a correlated pair which are subsequently ejected from the nucleus via one body currents (SRC or TC), can provide information on the correlations between them \cite{q_d_model,q_d_model_calcs,Boffi_book}. In an interaction with the one body nuclear current the (virtual) photon is absorbed by one nucleon of a correlated pair which is knocked out of the nucleus. This leaves the residual nucleus in an excited state and subsequently the second nucleon of the correlated pair is likely to be emitted from the nucleus.

Additional processes have to be taken into account when studying
NN-knockout reactions. There are competing two-body mechanisms such
as meson exchange currents (MEC) and isobar currents (IC) that contribute
to the measured cross section. Final state interactions (FSI) between
the outgoing nucleons and the residual nucleus and between the two
outgoing nucleons also complicate the picture. The importance of the contribution to the cross section of each of these other processes depends strongly on the reaction channel and hence studies of the ($\gamma$,pp), ($\gamma$,pn),
(e,e$'$pp) and (e,e$'$pn) can greatly improve our understanding
of the separate contributing mechanisms.

While NN-knockout experiments with real photons are only sensitive
to transverse components of the interaction, virtual photons are sensitive
to both the longitudinal and transverse components \cite{Boffi_book}. The longitudinal
cross section is dominated by one-body currents and is therefore more
sensitive to NN-correlations. Electron scattering experiments therefore
have significant advantages for investigating these properties. In
contrast two-body currents like IC and MEC are strongest in the transverse
cross section and are therefore more suitably studied with real photon
beams \cite{Boffi_book}.

Different properties of the nuclear response to virtual photons can
be studied using (e,e$'$pp) and (e,e$'$pn) reactions. For the (e,e$'$pp) reaction SRC are expected to dominate the longitudinal response. For the (e,e$'$pn) reaction TC are expected to play a major role in the longitudinal
nuclear response whereas SRC make a relatively minor contribution
\cite{16o_eep_tensor,16o_eepn_giusti}. Moreover the results of \cite{16O_eepN} and \cite{16o_eepn_giusti} suggest that TC influence the cross section principally through the Delta current.

In general two-body currents are more important in pn-knockout reactions
than in pp-knockout \cite{2N_knockout}. The contribution from MEC
to the (e,e$'$pp) cross section is suppressed because the virtual
photon will not couple to neutral pions. This is not the case for
pn-knockout where MEC make a substantial contribution to the cross
section, particularly in the transverse part. IC can contribute substantially
to the reaction cross section for pn knockout, their importance
grows as the transferred photon energy approaches the Delta-resonance
region. Again their contribution is more important in the transverse
nuclear response.

FSI between the outgoing nucleons and the residual nucleus have often
been treated theoretically using an optical model. However, it has
been shown recently \cite{16o_eepp_FSI} that the mutual interaction
between the two emitted nucleons can also have a large effect on reaction
cross sections. The magnitude of the effects depends strongly on the
reaction channel and on the particular kinematics. For instance an
increase in the calculated $^{16}$O(e,e$'$pp) cross section by nearly
an order of magnitude \cite{16o_eepp_FSI} has been found in super-parallel kinematics, where 
the two protons are emitted anti-parallel to one another,
at large recoil momentum. In pn emission the effect is partially
cancelled by destructive interference with the pion seagull MEC term \cite{16o_eeNN_FSI}.

The first electron scattering study of NN-knockout was a measurement
of the $^{12}$C(e,e$'$pp) reaction at NIKHEF \cite{12c_eepp_nikhef2,12c_eepp_nikhef}.
This experiment measured two-proton knockout at an energy transfer
$\omega$ = 212 MeV and a three-momentum transfer $|$\textbf{q}$|$ = 270 MeV/c.
The measured missing-energy spectrum shows a signature for knockout
of proton pairs from (1p)$^{2}$, (1p,1s), and (1s)$^{2}$ states. The data were
compared with a direct-knockout calculation which includes one- and
two-body currents. This comparison shows that the measured cross section
for the knockout of a (1p)$^{2}$ pair can largely be attributed to SRC.

Measurements at NIKHEF of the $^{16}$O(e,e$'$pp) reaction \cite{16O_ee_pp_NIKHEF_1,16O_ee_pp_NIKHEF_1b}
using large solid-angle proton detectors were able to separate the
ground state in the residual $^{14}$C nucleus. Clear signatures of SRC
were observed in transitions to this state. A further NIKHEF
measurement \cite{16O_ee_pp_NIKHEF_2} also obtained experimental evidence
for SRC from the measured energy-transfer dependence of the\newline $^{16}$O(e,e$'$pp)$^{14}$C$_{g.s.}$ reaction. This earlier evidence of SRC has received further confirmation in
recent (e,e'p) experiments at JLab \cite{JLab_eep}.

Similar results from the $^{16}$O(e,e$'$pp) reaction were obtained
with the three-spectrometer setup at the Mainz Microtron MAMI \cite{16O_ee_pp_mami}.
These measurements were carried out in super-parallel kinematics and
 various low-lying states in $^{14}$C were populated. Comparison with theoretical
calculations \cite{16O_eepp} provided evidence that SRC affect nucleon pairs with small centre-of-mass
momentum in relative S-states.

The (e,e$'$pp) reaction has also been measured over a wide range
of kinematics in the $^{3}$He nucleus \cite{3He_eepp}. At a momentum transfer of
\textbf{q} = 375 MeV/c data were taken at transferred energies ranging from
170 to 290 MeV and at $\omega$ = 220 MeV, measurements were performed
at \textbf{q} = 305, 375, and 445 MeV/c. The data are compared with continuum-Faddeev
calculations, which indicate that at $\omega$ = 220 MeV and for neutron
momenta below 100 MeV/c, the cross section is dominated by direct
two-proton emission induced by a one-body hadronic current. At higher
neutron momenta deviations between the data and calculations are attributed
to additional contributions from isobar currents.

There have been several measurements of real-photon two-nucleon knockout
reactions on $^{16}$O \cite{16O_gamma_NN,16O_gamma_pn,16O_gamma_pn_canada}.
The first measurement \cite{16O_gamma_NN}, carried out at the Mainz
180 MeV electron Microtron MAMI-A, measured both reaction channels
and covered the photon energy range 80-131 MeV. The energy resolution
of 7 MeV was too poor to resolve the different excited states within
the residual $^{14}$N nucleus. However, it was observed that the
strength of the ($\gamma$,pp) cross section for the emission of (1p)$^{2}$
nucleon pairs was very small and amounted to only $\approx$ 2\,\% of
the corresponding ($\gamma$,pn) cross section.

The study of ref \cite{16O_gamma_pn} measured the $^{16}$O($\gamma$,pn)
reaction at E$_{\gamma}$ = 72 MeV. An energy resolution of better
than 2 MeV was achieved which allowed cross sections for the excitation
of individual excited states in $^{14}$N to be measured. The ground (1$^{+}$),
3.95 MeV (1$^{+}$) and 7.03 MeV (2$^{+}$) states were all observed,
but there was no excitation of the 2.3 MeV (0$^{+}$) state.
Comparison was made with theoretical calculations including MEC and
IC. The calculated cross sections significantly underestimated the
strength of all the observed excited states but did predict the 3.95
MeV (1$^{+}$) state would be strongly excited, as observed. It was
deduced that states with CM pair orbital angular momenta L = 0 and
L = 2 both contributed strongly to the observed spectrum, and the absence
of the 2.3 MeV (0$^{+}$) state was attributed to the dominance
of interactions with $^{3}$S$_{1}$ pn pairs as assumed in the basic
quasi-deuteron process.

A further study of the $^{16}$O($\gamma$,pn) reaction \cite{16O_gamma_pn_canada}
 was carried out for the photon energy range E$_{\gamma}$
= 98.5 to 141.0\,MeV with an energy resolution of 2.8 MeV. Protons
were detected at two angles: 76$^{\circ}$ and 82$^{\circ}$ with
coincident neutrons being detected at corresponding quasi-deuteron
angles. Only the 3.95 MeV (1$^{+}$) state in the residual $^{14}$N
nucleus was significantly populated. This state is expected to have
a large L = 0 component compared to the other two states seen in the previous
experiment \cite{16O_gamma_pn} which both have strong L = 2 components.
The low recoil momentum range sampled in \cite{16O_gamma_pn_canada}
therefore favoured excitation of the 3.95 MeV (1$^{+}$) state.

In the present paper we report the results of the first measurements of the $^{16}$O(e,e$'$pn)$^{14}$N
reaction. The experiment was performed at the University of Mainz
855 MeV electron microtron MAMI. The data were centred on super-parallel
kinematics similar to those used in a previous $^{16}$O(e,e$'$pp)
measurement \cite{16O_ee_pp_mami}, also carried out at MAMI.

\section{Theoretical Framework}

In exclusive two-nucleon knockout electron scattering experiments
energy and momentum are transferred to a nucleus by a virtual photon
and the momenta of the scattered electron and both nucleons are determined.
Only processes where the recoil nucleus is left intact and no secondary
particles are created are considered here.

\subsection{Kinematics of the reaction}

\begin{figure}
\resizebox{0.90\columnwidth}{!}{
  \includegraphics{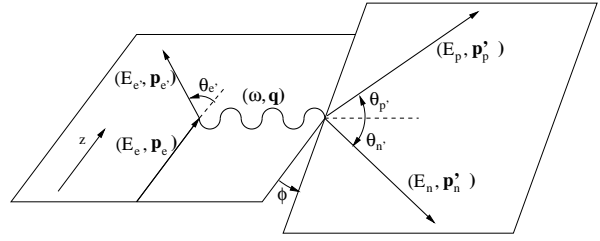}
}
\caption{Schematic of the kinematics for a two-nucleon knockout reaction.}
\label{fig_kinematics}
\end{figure}

The kinematics for the $^{16}$O(e,e$'$pn)$^{14}$N reaction are
shown schematically in figure \ref{fig_kinematics}. In the one photon
exchange approximation, the transferred virtual photon has an energy
$\omega=E_{e}-E_{e'}$ and three-momentum \textbf{q} = \textbf{p$_{e}$}
- \textbf{p$_{e'}$} where the subscripts e and e$'$ represent the
incident and scattered electron respectively. The proton and neutron
are ejected from the $^{16}$O nucleus with
momenta $\mathbf{p}_{p}^{'}$ and $\mathbf{p}_{n}^{'}$ and are detected
in coincidence with the scattered electron. In the present experiment
the data were centred on a super-parallel kinematic setting where
the proton is emitted in the forward direction parallel to the direction
of \textbf{q} and the neutron is emitted anti-parallel to \textbf{q}.

In an exclusive $^{16}$O(e,e$'$pn)$^{14}$N reaction the final
state can be completely reconstructed using the missing momentum

\begin{equation}
\mathbf{p}_{m}=\mathbf{q}-\mathbf{p}_{p}^{'}-\mathbf{p}_{n}^{'}=\mathbf{p}_{r}\label{eq:pm}\end{equation}
which is equal to the momentum of the undetected $^{14}$N recoil
nucleus \textbf{p$_{r}$}. Using a nonrelativistic approximation, the missing energy ($E_{m}$) can be defined as 

\begin{equation}
E_{m}=\omega-T_{p}-T_{n}-T_{r}=S_{pn}+E_{x}\label{eq:em}\end{equation}
where $T_{p}$, $T_{n}$ and $T_{r}$ are the kinetic energies of
the proton, neutron and recoil nucleus respectively; $T_{r}$ can
be determined from \textbf{p$_{r}$}. The $^{14}$N nucleus can be
left in a variety of states with excitation energy, $E_{x}$. $S_{pn}$
is the separation energy for the reaction.

The $^{2}$H(e,e$'$pn) reaction, used for calibration purposes, is
over-determined because both emitted particles were detected in the
final state and there is no residual nucleus. The missing energy is defined as 

\begin{equation}
E_{m}=\omega-T_{p}-T_{n}\label{eq:em2}\end{equation}
in this case.

\subsection{Theoretical Calculations}
\label{theory_calcs}

To help interpret the experimental data we compare the results to
the microscopic calculations of Giusti \emph{et al.} described in
refs \cite{16o_eepn_giusti,16o_eepp_giusti} with improvements to the
model described in ref. \cite{16O_eepN}. These were performed for
a number of kinematical settings covering the energy ranges and angular
ranges subtended by the experimental set-up and the average cross section
was used for comparison to the data. Table \ref{calc tab} shows the 
settings for which the calculations were performed.

\begin{table}
\begin{center}\begin{tabular}{|c|c|c|c|c|c|}
\hline 
Setting&
$\omega$ {[}MeV{]}&
\emph{q} {[}MeV/c{]}&
$\theta_{e'}$ {[}$^{\circ}${]}&
$\theta_{pq}$ {[}$^{\circ}${]}&
$\theta_{pn}$ {[}$^{\circ}${]}\tabularnewline
\hline
\hline 
1&
245&
333&
18&
6&
165\tabularnewline
\hline 
2&
185&
300&
18&
6&
165\tabularnewline
\hline
3&
215&
316&
18&
2&
165\tabularnewline
\hline 
4&
215&
316&
18&
8&
165\tabularnewline
\hline
5&
215&
316&
18&
6&
155\tabularnewline
\hline 
6&
215&
316&
18&
6&
175\tabularnewline
\hline
\end{tabular}\end{center}

\caption{\label{calc tab}The kinematical settings for which theoretical calculations
were performed. $\theta_{pq}$ and $\theta_{pn}$ are the angles subtended by the trajectories of the proton and virtual photon and the proton and neutron respectively.}
\end{table}

In the calculations the transition amplitude for the exclusive (e,e$'$pn)  knockout reaction contains contributions from both one-body and two-body hadronic currents. The one-body current contains the longitudinal charge term and the transverse convection and spin terms. The two-body current is derived performing a non relativistic reduction of the lowest-order Feynman diagrams with one-pion exchange. Therefore currents corresponding to the seagull and pion-in-flight diagrams and to the diagrams with intermediate Delta-isobar configurations are included \cite{16o_eepn_giusti,16o_eepp_giusti}.

The final state wave function includes the interaction of each one of the two outgoing nucleons with the residual nucleus, that is treated with a complex phenomenological optical potential containing central, Coulomb and spin-orbit terms. The mutual interaction of the two ejected nucleons is not included in the calculations. Although the effect of NN-FSI is non negligible for the (e,e$'$pn) cross section, quantitative effects are small and the qualitative features of the theoretical results are basically the same \cite{16o_eeNN_FSI}.

In the initial state the two-nucleon overlap function between the ground state of the target and the final state of the residual nucleus has been computed partitioning the Hilbert space, in order to determine the contribution of long-range correlation (LRC) and SRC separately.  The LRC, describing the collective motion at low energy as well as the long-range part of TC, were computed using the self consistent Green's function (SCGF) formalism \cite{NN_int_rev} in an appropriate harmonic oscillator basis. The effects of SRC due to the central and tensor part at high momenta were added by computing the appropriate defect functions. A crucial contribution is given here by TC, which  produce defect functions also for channels for which the uncorrelated wave function vanishes. This partitioning process is justified by the separation of the momentum scales associated to SRC and LRC. The Bonn-C interaction \cite{Bonn_C} was used in the calculations.

\section{Experimental Set-Up}

The measurements were performed using the electron scattering facility
(3-spectrometer facility) of the 100\,\% duty factor Mainz Microtron MAMI \cite{MAMI_Albequrqe}.
The 855 MeV, 10 to 20 $\mu$A electron beam was incident on a waterfall
target \cite{Waterfall_target} of thickness of 74 mg cm$^{-2}$.
A deuterium target \cite{SimonSirca,Cryo_target_web} was used for
detector calibration. This target set-up consisted of a cylindrical
cell containing liquid $^{2}$H with a nominal thickness of 165 mg cm$^{-2}$.
Beam currents of 0.25, 0.5 and 1 $\mu$A were used for calibration.

\begin{figure}
\resizebox{0.90\columnwidth}{!}{
  \includegraphics{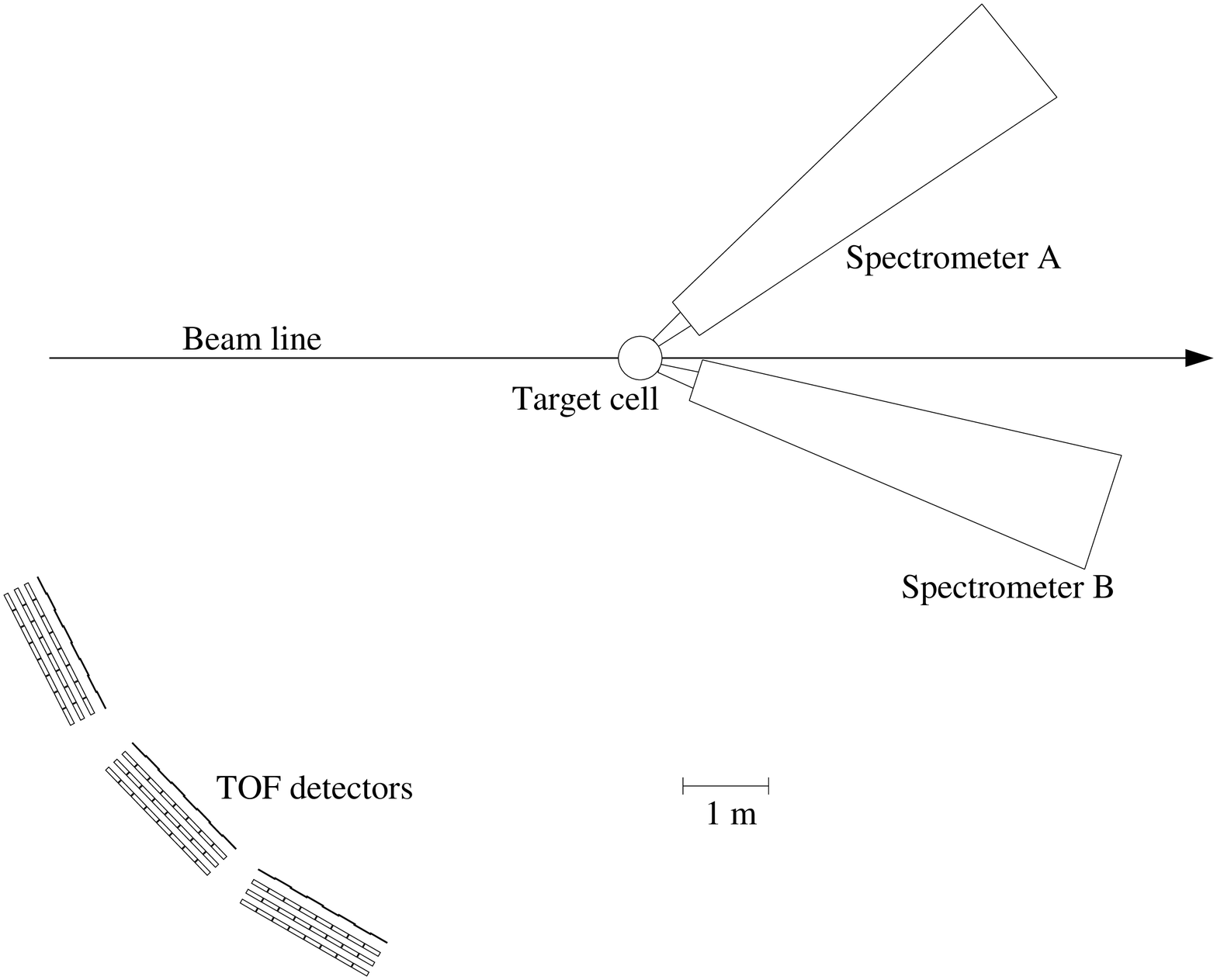}
}
\caption{Diagram of the detector set-up for the $^{16}$O(e,e$'$pn) experiment using the waterfall target.}
\label{fig_16o_setup}
\end{figure}

\begin{figure}
\resizebox{0.90\columnwidth}{!}{
  \includegraphics{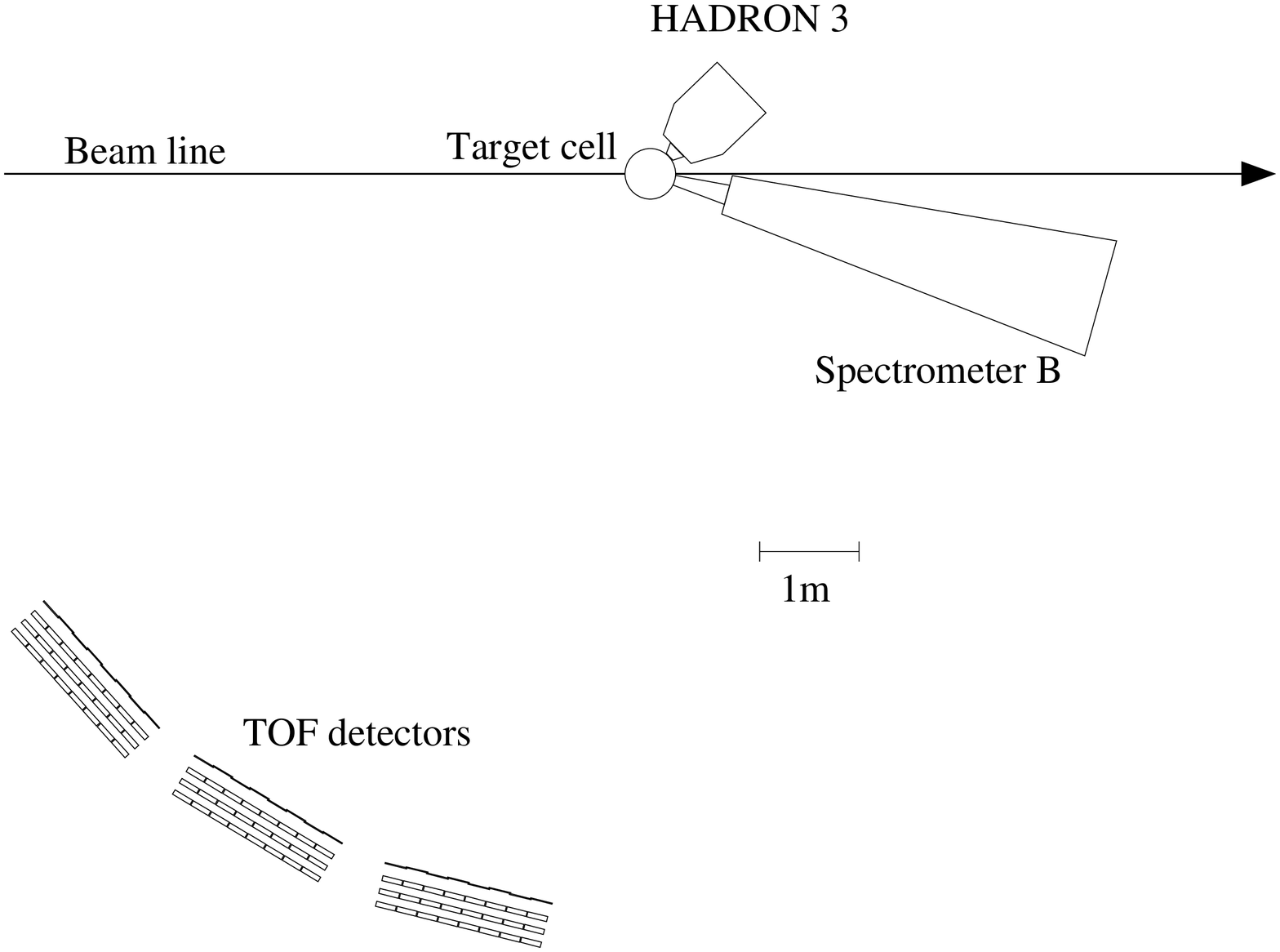}
}
\caption{Diagram of the detector set-up for the calibration
runs using the $^{2}$H cryogenic target.}
\label{fig_3he}
\end{figure}

For both targets the scattered electrons were detected in Spectrometer
B \cite{SpecB} which was placed at forward angles with respect to
the beam, see table \ref{kine_tab} and figures \ref{fig_16o_setup}
and \ref{fig_3he}. Spectrometer B is a magnetic spectrometer which
has a solid angle of $\Delta\Omega$ = 5.6 msr and momentum acceptance
of $\Delta p/p$ = 15\,\%, momentum resolution of $\delta p/p$ $\leq$
10$^{-4}$ and angular resolution of $\delta\theta$ = $\delta\phi$
= 1.5 mrad.

\begin{center}%
\begin{table*}
\begin{center}\begin{tabular}{|c|c|c|c|c|c|c|c|}
\hline 
Target&
$\omega$ {[}MeV{]}&
\emph{q} {[}MeV/c{]}&
$\theta_{e'}$ {[}$^{\circ}${]}&
$\theta_{p}$ {[}$^{\circ}${]}&
$\theta_{n1}$ {[}$^{\circ}${]}&
$\theta_{n2}$ {[}$^{\circ}${]}&
$\theta_{n3}$ {[}$^{\circ}${]}\tabularnewline
\hline
\hline 
$^{16}$O&
215&
316&
18&
-38.8&
120&
136.5&
155.5\tabularnewline
\hline 
$^{2}$H&
190&
300&
15.5&
-44&
97&
117&
137\tabularnewline
\hline
\end{tabular}\end{center}

\caption{\label{kine_tab}Overview of the kinematic configurations for each
target. The subscript e', p and n(1,2,3) represent the central position in the experimental hall of the electron and proton detectors and the three neutron detector stands respectively. The incident beam energy was 855 MeV.}
\end{table*}
\end{center}

To detect protons Spectrometer A \cite{SpecB} was used with the $^{16}$O
target and the HADRON3 (H3) detector \cite{HADRON3} from NIKHEF with
the $^{2}$H target. Again these were both placed at forward angles
with respect to the beam but on the opposite side of Spectrometer
B, thus parallel to \textbf{q}. Spectrometer A is a large acceptance magnetic spectrometer with
solid angle $\Delta\Omega$ = 28\,msr and momentum acceptance $\Delta p/p$
= 20\,\%; the momentum and angular resolutions are the same as those
of Spectrometer B. H3 is a large solid angle ($\Delta\Omega$ = 230\,msr) hodoscope consisting of 128 bars of plastic scintillator divided
into eight layers; two hodoscope layers and six energy determining
layers. H3's proton energy acceptance range is 70 - 225 MeV. The energy
resolution of H3 is 2.7\,\% and the angular resolution is $\delta\theta$
= 0.52$^{\circ}$ and $\delta\phi$ = 1.04$^{\circ}$.

The Glasgow-T\"{u}bingen time-of-flight (TOF) detector system \cite{TOF}
was used for detection of neutrons for both targets. The TOF detector
system consists of 96 bars of plastic scintillator, 72 of which are
5 cm thick 'TOF' bars for neutron detection behind 24 'Veto'
detectors, each 1 cm thick, used to discriminate between charged and neutral particles.
The bars were arranged in three separate stands which were placed
at backward angles with respect to the electron beam on the same side
as Spectrometer B and so anti-parallel to \textbf{q}. Each stand consisted of 3 layers of 8 TOF bars
which were positioned behind a layer of 8 overlapping Veto detectors.
The position and timing resolutions of the TOF bars (FWHM) are $\leq$
6 cm and $\leq$ 0.4 ns respectively. The TOF stands were positioned at a distance
of $\approx$ 7.5 m from the target giving angular resolutions of
$\delta\theta$ = 1.5$^{\circ}$ and $\delta\phi$ = 0.5$^{\circ}$.
For the detected neutron energy range of 30 to roughly 65\,MeV an average
energy resolution of $\approx$ 2 MeV (FWHM) was achieved. The
neutron energy threshold was set in the analysis software at 30 MeV
 (see section \ref{sec:neutron_efficiency}).

To detect reactions involving the particles of interest coincidence-electronics were used. Each of
the three detectors involved (Spectrometer B, H3/Spectrometer A and
TOF) generated a signal on the detection of a particle meeting detector-specific
requirements. In the set-up used here only an electron-proton coincidence
was required to trigger an event read-out; the TOF detector systems
were started when an electron was detected in Spectrometer B. TDCs
provided the relative time information which was used to determine
whether an event was in the threefold or twofold coincidence region
or was completely random (see section \ref{sec:Random-subtraction}).

\section{Analysis}

Using established procedures \cite{SpecB,HADRON3} the momenta of
the scattered electron and emitted proton were determined. The energy
of the emitted neutron was determined from its time-of-flight as described
below. For each detector used in the experiment the arrival
time of trigger signals was measured when a particle was detected and later corrected
offline for time-of-flight differences and detector specific delays.
This allowed real event coincidence times to be determined and prompt
and random regions to be defined for subtraction of random coincidence
events.

\subsection{\textbf{$t_0$} determination}

To determine the energy and momentum of the emitted neutron an accurate
measurement of the flight-time from the reaction vertex to the position
of detection within the TOF detector is required. The uncalibrated
measured flight-time contains contributions from delays caused by
signal processing in the TOF detector. The effect of all such delays
is referred to as the '$t_0$' value which must be subtracted to determine
real neutron flight times.

In the experiment the start time for the TOF detectors was given by
Spectrometer B. Corrections for delays in Spectrometer B were made
so that the start-time was that from when the reaction took place
within the target cell. This was possible since the particle flight-time calibrations and offsets in Spectrometer B had been determined
previously \cite{SpecB}. Any further delays were then only from the
TOF system. The $t_0$ value for each TOF bar was determined using
data from the highly overdetermined $^{2}$H(e,e$'$pn) reaction. For each event the energy and momentum
of the virtual photon were determined from the known incident beam
energy and the angle and momentum of the scattered electron measured
in Spectrometer B. Then using the measured polar and azimuthal angles
of the neutron in TOF together with conservation of energy and momentum
it is possible to evaluate the
neutron energy and expected time of flight. When this expected flight
time was subtracted from the measured flight time in TOF a constant
offset, the $t_0$, was left for all real neutron events while random
coincidences were spread over the measured time range. The mean
values of the peaks were determined to obtain $t_0$ values for each
of the TOF bars.

\subsection{Determination of the neutron detection efficiency\label{sec:neutron_efficiency}}

While the detection efficiency of the electron and proton spectrometers
is close to unity \cite{SpecB} the detection efficiency for neutrons in plastic scintillator is much lower, being $\approx$ 1\,\%/cm.

The method employed to determine the neutron detection efficiency
was by use of a Monte Carlo simulation developed by Stanton \cite{stanton}
and later improved by Cecil \emph{et al.} \cite{Cecil}. This code
calculates the neutron detection efficiency as a function of neutron
kinetic energy and the detector threshold and a given detector material and geometry. 
Known cross sections for
most of the possible neutron reaction channels are included to give
an overall reaction cross section for a neutron traversing the scintillator
material. The output of the code has been tested against
measurements of the neutron detection efficiency for various types of
plastic scintillator and detector dimensions and was found to be accurate
to within about 10\,\% \cite{Cecil}. 

A global software threshold was applied to the experimental data and
the value of this threshold was applied to the Stanton code. The threshold
energy was calibrated using proton energy deposition from consecutive TOF layers \cite{TOF}, and
applied to the geometric mean of the pulse height recorded at both
ends of the TOF bars. It was set above the effective hardware threshold
so that it was independent of position. The threshold used was 20
MeV$_{ee}$ $\equiv$ 30 MeV neutron energy.

To model the efficiency of stands of three TOF layers, the detection
efficiency for a bar in the second TOF layer was approximated by $\rm\varepsilon\phi_r$
where $\varepsilon$ is the detection efficiency for one bar and $\rm\phi_r$ is equal to $(1-\varepsilon)$
which approximately accounts for the reduction in neutron flux reaching
the second layer. Similarly a bar in the third layer is taken to have
efficiency $\rm\varepsilon\phi_r^{2}$. The overall detection efficiency for
the three layers was then the sum of the efficiencies for each individual
layer. Any broken TOF bars in a set of three were not included in
the overall efficiency although their effect on the neutron flux was
still included.

To test the validity of the model of the neutron detection efficiency described above a comparison was made between the modelled efficiency using the Stanton code and the experimentally measured efficiency \cite{duncan}. 

\begin{figure}
\resizebox{1\columnwidth}{!}{
  \includegraphics{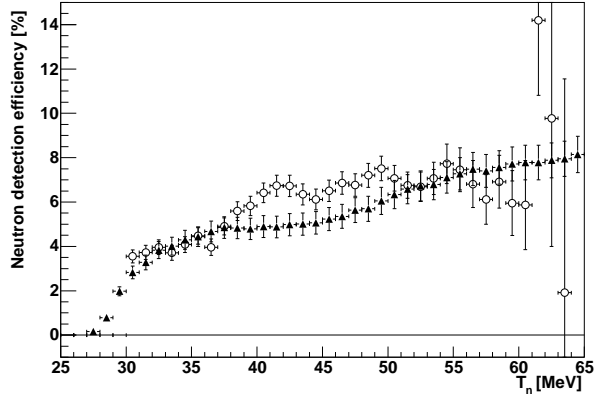}
}
\caption{Experimental neutron detection efficiency
in the TOF detector system (circles) compared to that predicted by the Stanton
code (triangles) \cite{Cecil}.}
\label{fig_stanton_comparison}
\end{figure}

To do this the ratio of yield from the $^{2}$H(e,e$'$pn) reaction
to the yield from the $^{2}$H(e,e$'$p)n reaction, where the recoil
neutron would have been above threshold in TOF, was determined. Data from
all the separate TOF bars have been added to obtain this ratio. The
comparison between this measured neutron detection efficiency in TOF
and that predicted by the Stanton code model is shown in figure
\ref{fig_stanton_comparison}. 

The modelled efficiency was found to show fairly good agreement to
the measured efficiency for neutron energies of 30 - 40 MeV but under-predicted
the measurements by an average of $\approx$ 20\,\% for neutrons in 
the range 40 - 50\,MeV. The deuterium data used for this comparison were taken when H3 was used for detection of the proton. It is believed that the structure seen in the measured efficiency between 40 and 50 MeV is due, in part, to the layering of the H3 detector. At the boundaries between scintillator layers the energy of some events is incorrectly determined because of pulsheight thresholds.

With the $^{16}$O set-up the range of detected neutron energies was greater than that for the $^{2}$H target, 30 to $\leq$ 130 MeV, so the efficiency model of three TOF bars employing the Stanton code was used. For the range of neutron energies where it was possible to make a reasonable comparison, 30
- 60 MeV, it was found that the Stanton code under-predicted the measured
data by an average of $(13\pm2)$\,\%. Therefore the efficiency taken from the Stanton code was increased universally by 13\,\% for the evaluation of the cross sections.

\subsection{Random subtraction\label{sec:Random-subtraction}}

\begin{figure}
\resizebox{1\columnwidth}{!}{
  \includegraphics{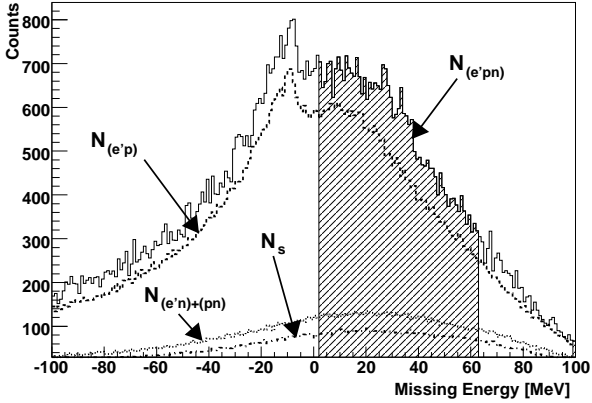}
}
\caption{
Missing energy plot showing the contributions from
the different types of random coincidences. The solid line, labelled $N_{e'pn}$, represents
the observed triple coincidence events; the dashed line, labelled $N_{e'p}$, represents
random neutron events; the dotted line, labelled $N_{(e'n)+(pn)}$, represents random proton and
random electron events and the dot-dashed, labelled $N_{s}$, line represents events where
all three detected particles are in random coincidence.
 The hatched
area shows the $E_{m}$ region covered in the excitation energy spectra
of figures \ref{14N_ex_pic} and \ref{14n_ex_pm_cut1.pic}.}
\label{fig_em_bg}
\end{figure}

\begin{figure}
\resizebox{1\columnwidth}{!}{
  \includegraphics{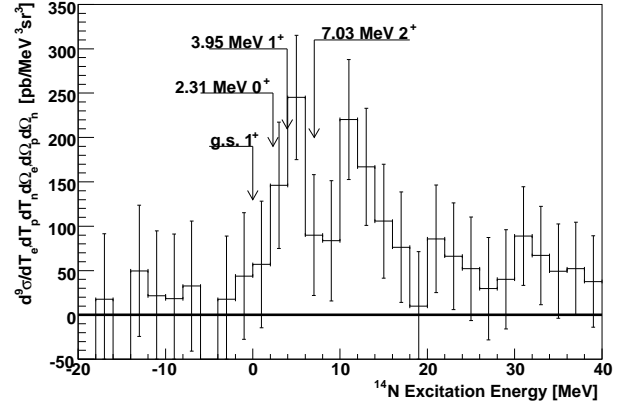}
}
\caption{Excitation energy of the residual $^{14}$N nucleus
after correction for random coincidences. The positions of the four
lowest states in the $^{14}$N are shown.}
\label{14N_ex_pic}
\end{figure}

In a triple coincidence experiment besides true triple coincidence events, various types of random coincidence events are also detected. These random events have to be subtracted from the experimental yield. The missing energy plot for this
experiment before the subtraction of random events is shown in figure
\ref{fig_em_bg} which also shows the contributions due to the different
types of random coincidence events.
The solid line, labelled $N_{e'pn}$, shows the observed
triple coincidence events, the dashed line, labelled $N_{e'p}$, shows the contribution
from true electron-proton coincidences with a random neutron, the dotted
line labelled, $N_{(e'n)+(pn)}$, shows the contribution from events with true electron-neutron
or true proton-neutron coincidences with the third particle being random and
the dot-dashed line, labelled $N_{s}$, is from events where all three particles are in
random coincidence.
 The hatched area indicates the $E_{m}$ region covered in the excitation energy spectra of figures \ref{14N_ex_pic} and \ref{14n_ex_pm_cut1.pic}.

As can clearly be seen in figure \ref{fig_em_bg} the largest contribution
of random events was from those associated with a random neutron in
the TOF detector. With the trigger used,
when there was an acceptable electron-proton event, the TOF detectors
 were read out giving rise to the large number of random
events covering the time window sampled. 

The contributions from the different types of random events are determined
from analysing data in different regions on a $t_{e-p}$ vs $t_{n}$
plot where $t_{e-p}$ is the time difference between electron and
proton detection and $t_{n}$ is the neutron flight time. The arrival-time distribution of random neutrons falls exponentially with increasing flight-time \cite{neutron_bg}.
This is because, with the QDC gates being controlled by the electron
detector trigger, a particle from a nuclear reaction other than the
one that triggered the electron detector can arrive earlier in time
than the corresponding neutron. This particle can 'steal' the QDC gate
and the information pertaining to this second particle is processed
by the detector instead. The chance of this happening falls exponentially
with neutron flight time. To account for this a correction factor
was applied to the random neutron events when performing the background
subtraction: an exponential function was fitted to a region in the
neutron flight time spectrum where only random events were expected
and this was extended over the prompt-event region \cite{duncan}. The ratio of number
of events in the prompt region, determined using the fit, to that in the random region was then used as the correction factor. The factor $f_{c}$ used was 1.037.

Having selected the regions to be used in the subtraction the number
of prompt coincidences was determined using

\begin{equation}
True_{(e'pn)}=N_{(e'pn)}-f_{c}N_{(e'p)}-N_{(e'n)+(pn)}+f_{c}N_{s}\label{bg_corr_eq}\end{equation}
where the subscript terms in brackets represent the different types
of coincidence-event regions and the subscript \emph{s} represents
the threefold uncorrelated events; $f_{c}$ is the correction factor
applied to the random neutron events as described above. The addition of $f_{c}N_{s}$ corrects for the 'extra' subtraction of 3-fold randoms included in the $f_{c}N_{(e'p)}$ term.
 In the (e$'$p) and (s) samples the neutron
flight times were reassigned to be within the prompt neutron time
region in order to determine the correct neutron energies. The time ranges
covered by the TDCs was insufficient to obtain separate samples of
the (e$'$n) and (pn) regions so these two contributions were treated
as if they were all (e$'$n) events. Analysing the joint (e$'$n)
and (pn) as (pn) events was also tried but this made no appreciable
difference to the final results, the large random neutron contribution
dominating the subtraction.

After the subtraction procedure is carried out real\newline (e,e$'$pn) coincidence events are left, as shown in the $^{14}$N excitation spectrum of figure \ref{14N_ex_pic}. The peak at $\approx$ 4 MeV corresponds to the position expected for the 3.95 MeV (1$^{+}$) state. A range of other states are observed in $^{14}$N at E$_{x}$ $\geq$ 9 MeV. This is discussed further in the results section

\subsection{Determination of cross sections}

The cross section for a given kinematic
variable $X$ (e.g. $p_{m}$) is determined in the following
way:

\begin{equation}
\frac{d^{8}\sigma}{dV^{8}}(X)=\int_{E_{x}}\frac{N(X)}{\int\mathcal{L}dt\mathcal{V}(X)}\left|\frac{\delta T_{2}}{\delta E_{x}}\right|dE_{x}\label{eq:x-sec}\end{equation}
where $N(X)$ represents the number of true (e,e$'$pn) events
for a given excitation energy range, $\int\mathcal{L}dt$ is the integrated
luminosity and $\mathcal{V}(X)$ is the experimental detection
volume in phase space. The factor $\left|\delta T_{2}/\delta E_{x}\right|$ is
a Jacobian where $T_{2}$, in this case, is the neutron kinetic energy and $E_{x}$ is the missing energy range over which events are integrated.

The detection volume is calculated using a Monte-Carlo method with a nine-dimensional volume $\mathcal{V}$ \cite{Aqua}. It
takes into account the energy and angular acceptances of each of the
detectors involved in the experiment. The neutron detection efficiency
of the TOF detector system was included as a weight for events generated
with the Monte-Carlo.

In electron scattering experiments the electron of interest can
radiate photons reducing the energy of the incoming or scattered
electron. This is usually evident in reconstructed missing energy
spectra by the presence of a radiative tail, an example of which can
be seen in figure \ref{pic:2H_Em}. The Monte-Carlo program used to
determine the detection volume includes multi-step radiative corrections
following the formul$\ae$  of Mo and Tsai \cite{radiative_corrections}
to account for this.

\subsection{Energy Resolution}

The energy resolution of the apparatus determined from calibration $^{2}$H(e,e$'$pn) data taken when H3 was used for detection
of the emitted proton. As the energy resolution
of Spectrometer A is better than that of H3 this procedure only provides
an upper limit of the energy resolution for the $^{16}$O(e,e$'$pn)
experiment.

\begin{figure}
\resizebox{1\columnwidth}{!}{
  \includegraphics{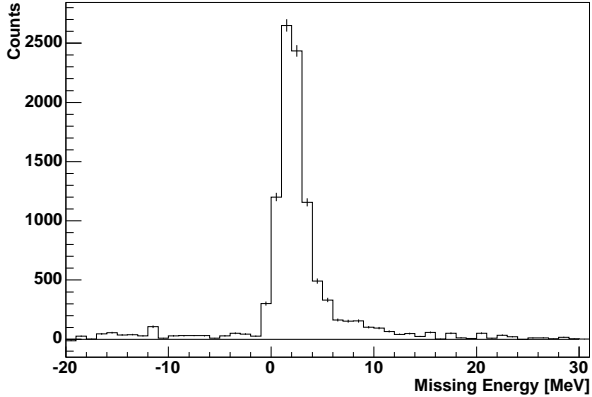}
}
\caption{Missing energy distribution for the $^{2}$H(e,e$'$pn) reaction.
The peak has a mean of 2.1 MeV and the resolution amounts to 3.0 MeV
(FWHM).}
\label{pic:2H_Em}
\end{figure}

The missing energy peak for the $^{2}$H(e,e$'$pn) reaction is shown
in figure \ref{pic:2H_Em} after correction for random background
contributions. The mean of the peak was determined and has a value of 2.1 MeV with a FWHM value
of 3.0 MeV. This must be regarded as an upper limit for the resolution of the $^{16}$O(e,e$'$pn) experiment.

\section{Results}

Figure \ref{14N_ex_pic} shows the background-corrected excitation-energy spectrum of the $^{16}$O(e,e$'$pn)$^{14}$N reaction. The positions
of the ground state and the first three low lying excited states in
$^{14}$N are marked. There is a prominent peak around the energy
expected for the 3.95 MeV (1$^{+}$) state in $^{14}$N. Given the resolution
of the experiment ($\leq$ 3.0 MeV FWHM) this peak will also
contain contributions from the 2.31 MeV (0$^{+}$) and 7.03 MeV (2$^{+}$)
excited states if they have been populated. The $^{14}$N ground state appears to be at most rather
weakly excited.

There is a second strong peak centred at $\approx$ 11 MeV which corresponds
to states in the continuum region in the residual $^{14}$N nucleus. The width
of this peak is such that several states probably contribute and it
is not possible to make a reasonable attribution of this peak to any
particular states.

The real photon studies of the $^{16}$O($\gamma$,pn) reaction
\cite{16O_gamma_pn,16O_gamma_pn_canada} also both observed strong
excitation of the 3.95 MeV (1$^{+}$) state. The study of Isaksson
\emph{et al}. \cite{16O_gamma_pn} observed strong excitation
of the ground state (1$^{+}$) and the 7.03 MeV (2$^{+}$) state, but
neither observed the 2.31 MeV (0$^{+}$) state. The photon energies
used in both these experiments were somewhat lower than the energy
transfer of the present work.

\begin{figure}
\resizebox{1\columnwidth}{!}{
  \includegraphics{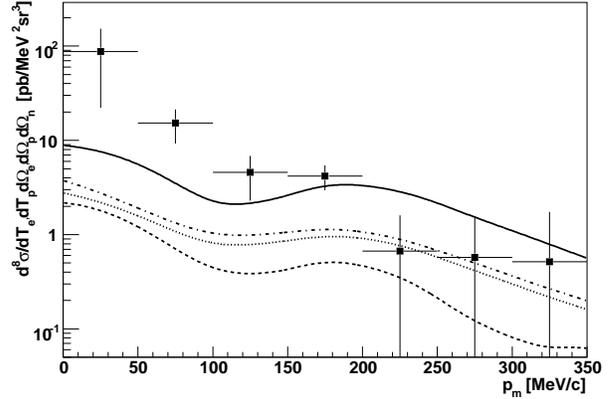}
}
\caption{Cross section for events in the range 2 $\leq$
E$_{x}$ $\leq$ 9 MeV compared to calculations of the Pavia group.
Calculations for transitions to the first three excited states, 3.95
MeV (1$^{+}$), 2.31 MeV (0$^{+}$) and 7.03 (2$^{+}$) are included
in the calculations. The dashed line shows the cross section for the
one body part of the reaction only; the dotted line also includes
the $\pi$-seagull term; the dashed dotted includes the one-body,
$\pi$-seagull term and pion-in-flight terms and the solid line is
for the complete cross-section including contributions from IC.}
\label{pav_x_sec_pic}
\end{figure}

\begin{figure}
\resizebox{1\columnwidth}{!}{
  \includegraphics{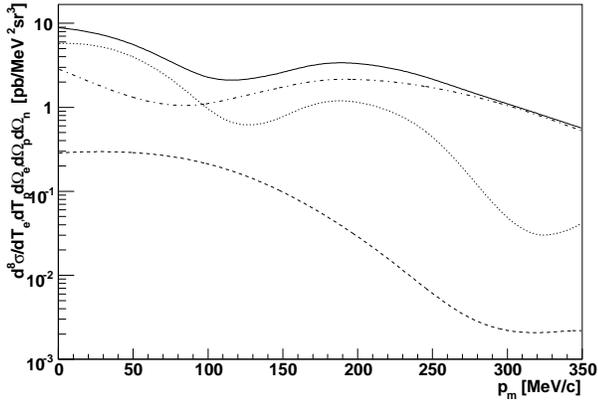}
}
\caption{Theoretical cross sections from the calculations of the Pavia group \cite{16O_eepN} for the 2.31 MeV (0$^{+}$), 3.95 MeV (1$^{+}$), 7.03 (2$^{+}$) and the three states combined; represented by the dashed, dotted, dashed-dotted and solid lines respectively. The plots are for the full cross section including the one body, $\pi$-seagull, pion-in-flight and IC terms.}
\label{theory_states_comp}
\end{figure}

Theoretical calculations of the $^{16}$O(e,e$'$pn) reaction for transitions to the lower lying states in $^{14}$N have been carried out by the Pavia group, see section \ref{theory_calcs} and figures \ref{pav_x_sec_pic} and \ref{theory_states_comp}. These calculations predict cross sections roughly an order of magnitude weaker than observed experimentally, though the relative strength of the different states in $^{14}$N is similar to what is seen.
 The calculations predicted that transitions to the 3.95 MeV (1$^{+}$) state would be the largest which is consistent with the observed missing energy spectrum. Transitions to the 7.03 MeV (2$^{+}$) were also predicted to be strong.

DW calculations of the $^{16}$O(e,e$'$pn) reaction, including the effects of MEC, IC, SRC and TC, have also been carried out by the Ghent group for the four low lying excited states in $^{14}$N marked in figure \ref{14N_ex_pic} \cite{16o_eepn_tensor}. The calculations were carried out in the same super-parallel kinematics as the measurement; again these calculations predict cross sections roughly an order of magnitude smaller than the data. Similarly to the Pavia calculations these predict that transitions to the 3.95 MeV (1$^{+}$) state are strongest; the occupation of the neighbouring excited states states is an order of magnitude smaller, which is the same trend as is seen in the data.

Figure \ref{pav_x_sec_pic} shows the measured differential cross
section, as a function of absolute magnitude of the missing momentum,
for the group of states from 2 to 9 MeV residual $^{14}$N excitation
energy. The spectrum has its maximum value at low missing momentum, falls rapidly to around 120 MeV/c before apparently levelling off and then falling more slowly. The experimental data are compared with the calculations from the Pavia group \cite{16O_eepN}, described in section \ref{theory_calcs}, which include the sum of contributions of transitions to the  2.31\,MeV (0$^{+}$), 3.95 MeV (1$^{+}$) and 7.03 MeV (2$^{+}$) states. Figure \ref{theory_states_comp} shows a comparison of the full cross sections for transitions to the three different excited states included in the curves of  figure \ref{pav_x_sec_pic}. The calculations significantly underestimate the measured cross section at low missing momenta. Above 100 MeV/c there is reasonable agreement with the data but the error bars in the data are large. Figure \ref{theory_states_comp} shows that below 100 MeV/c the main strength is predicted to be from transitions to the 3.95 MeV (1$^{+}$) state and above this to the 7.03 MeV (2$^{+}$) state.

Figure \ref{pav_x_sec_pic} also shows the cumulative contributions of the Pavia calculations from 1-body, seagull, pion-in-flight and isobar currents. The largest contribution to the cross section comes from the isobar currents and in the Pavia calculations the dominant contribution to these are from TCs. The large discrepancy between the experimental and calculated cross sections and the large statistical errors in the data do not allow strong conclusions to be drawn from the comparison.

The data can be compared with the super-parallel missing momentum
spectra observed in the $^{16}$O(e,e$'$pp) reaction \cite{16O_ee_pp_mami}.
This experiment excited the ground state (0$^{+}$) in the residual $^{14}$C
nucleus and excited states at 7.01 MeV (2$^{+}$), 8.32\,MeV (2$^{+}$), 9.75
MeV (0$^{+}$) and 11.3 MeV (1$^{+}$). The sum of all the contributions to the different states in the $^{14}$C is less than 1.4 pb MeV$^{-2}$sr$^{-3}$. In contrast the summed strength for the group of residual states in $^{14}$N for 2 $\leq$ E$_{x}$ $\leq$ 9 MeV is $\approx$ 2 orders of magnitude greater. This ratio is of a similar magnitude to the factor of around 50 seen in the $^{16}$O($\gamma$,pp) and ($\gamma$,pn) work of reference \cite{16O_gamma_NN}.

\begin{figure}
\resizebox{1\columnwidth}{!}{
  \includegraphics{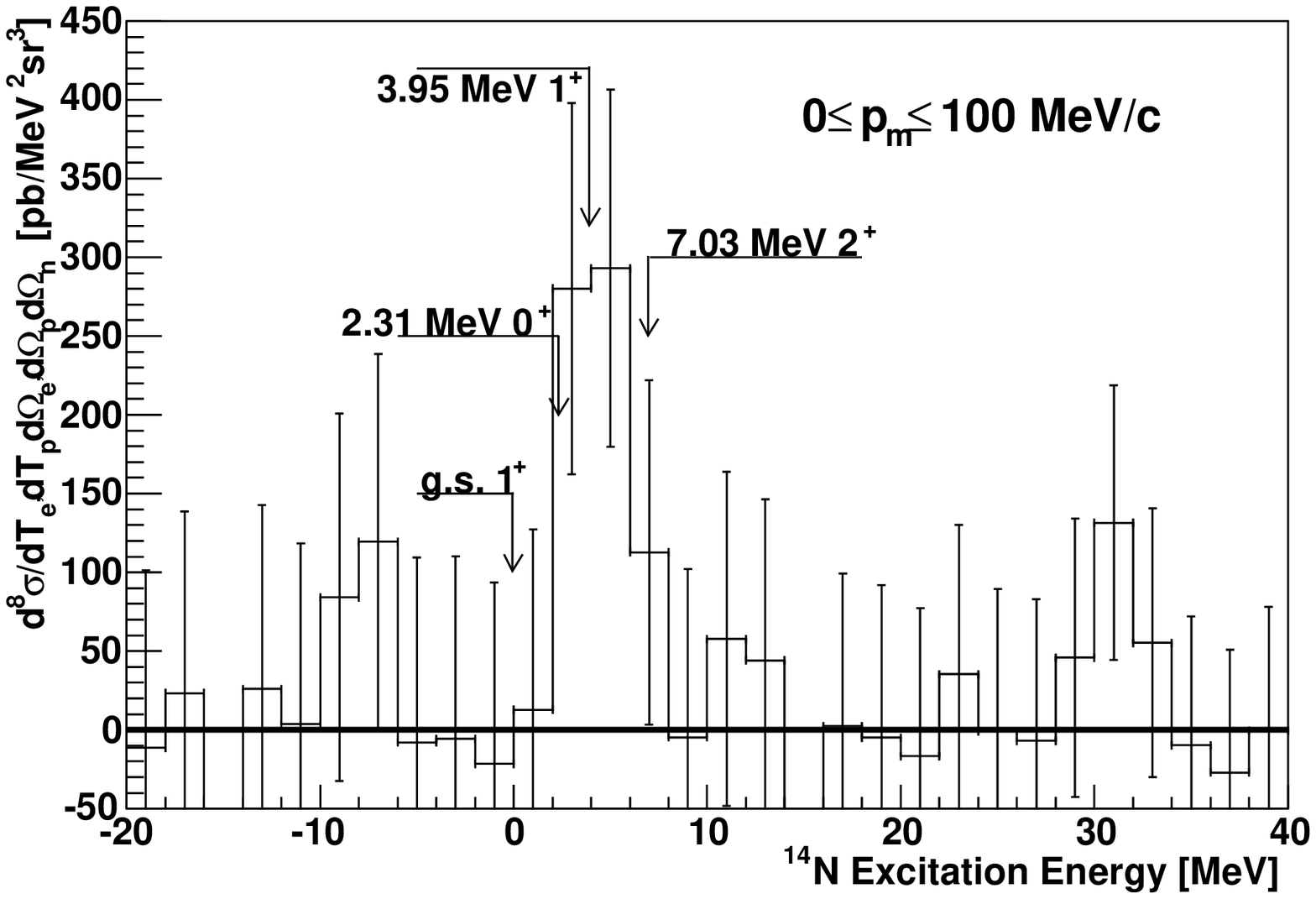}
}

\resizebox{1\columnwidth}{!}{
  \includegraphics{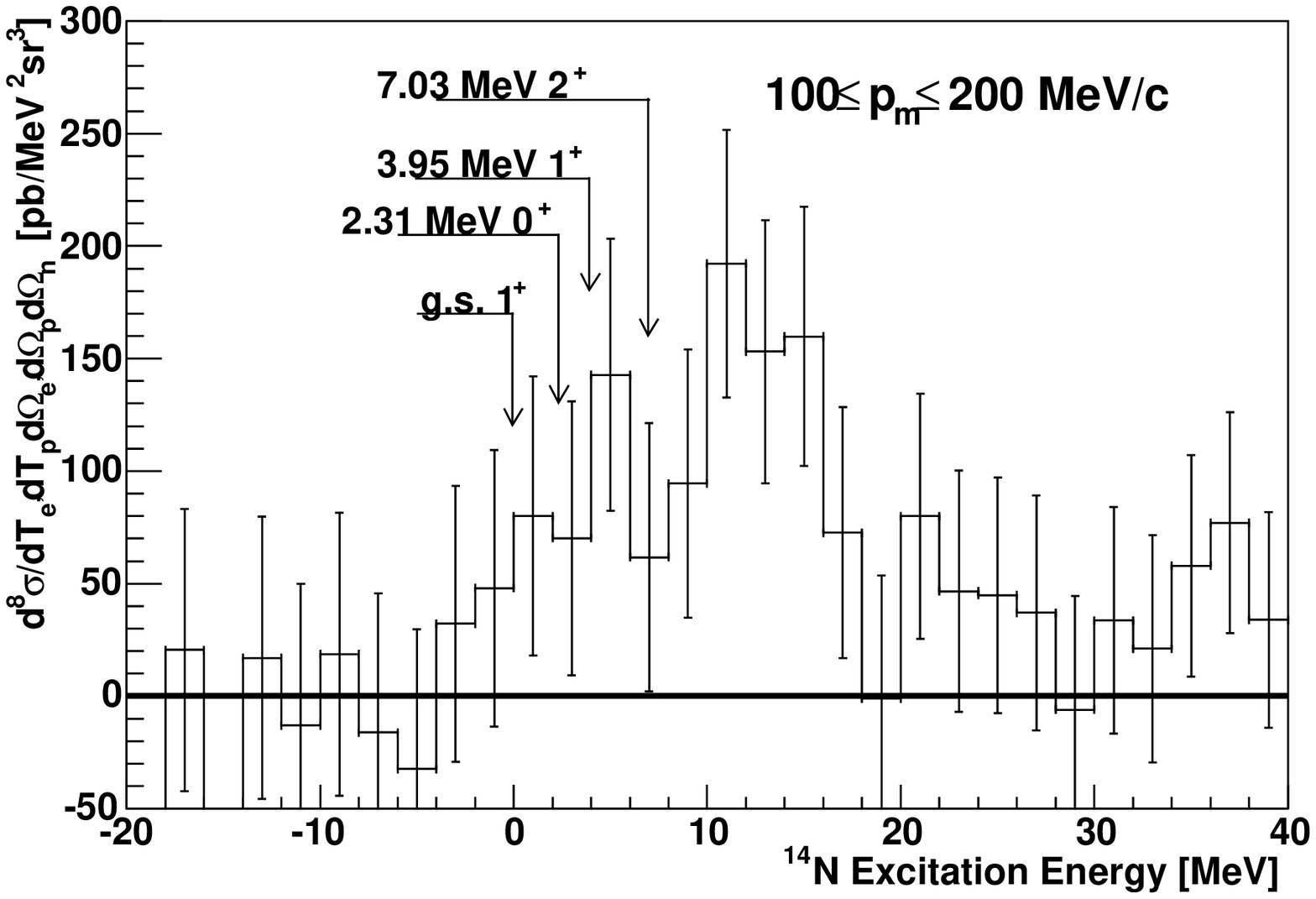}
}

\caption{Excitation energy of the residual $^{14}$N
nucleus. Figure (a) shows the excitation energy spectrum for the events
in the range 0$\leq$$p_{m}$$\leq$100 MeV/c while figure (b) shows
the excitation energy spectrum for the events in the range 100$\leq$$p_{m}$$\leq$200
MeV/c.}
\label{14n_ex_pm_cut1.pic}
\end{figure}

Figure \ref{14n_ex_pm_cut1.pic} shows the residual $^{14}$N excitation
spectrum cut into two regions of missing momentum: a) p$_{m}$ $\leq$ 100 MeV/c and b) 100 $\leq$ p$_{m}$ $\leq$ 200 MeV/c. The group of
states from 2 $\leq$ E$_{x}$ $\leq$ 9 MeV is present in both missing momentum regions. The excitation
energy distribution is somewhat broader and weaker in the higher missing
momentum region. This variation in strength supports the idea that
the strong 3.95 MeV (1$^{+}$) state has its maximum strength at low recoil momentum. Although the statistics are poor figure \ref{14n_ex_pm_cut1.pic} (b) also shows apparent strength
around 0 MeV which is absent in figure \ref{14n_ex_pm_cut1.pic} (a).
This may suggest that the $^{14}$N ground state has most of its strength in
the higher missing momentum region.

\begin{figure}
\resizebox{1\columnwidth}{!}{
  \includegraphics{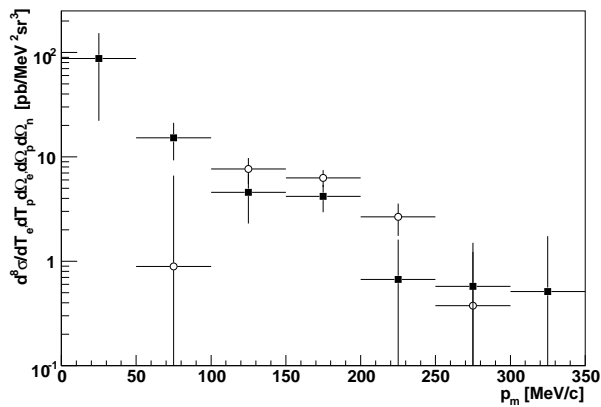}
}
\caption{Comparison of cross sections of events in
the excitation energy range 2 $\leq$E $_{x}$ $\leq$9 MeV (squares)
and 9 $\leq$E $_{x}$ $\leq$15 MeV (circles).}
\label{comp_x_sec_pic}
\end{figure}

 The group of higher energy residual states in $^{14}$N have nearly all of their strength in the higher missing momentum range which is similar to what was seen in ref \cite{12c_gamma_2N_2}. In that reference the behaviour was attributed to the dominant states having L $>$ 0. Figure \ref{comp_x_sec_pic}, which shows the missing momentum dependence of two groups of states, 2 $\leq$ E$_{x}$ $\leq$ 9 MeV and 9 $\leq$ E$_{x}$ $\leq$ 15 MeV, lends further support to this idea. Although for lower missing momentum it is difficult to accurately determine the shape of the cross section because of lack of data and poor statistics it appears that the higher energy group of states have no strength at zero missing momentum and their distribution peaks at around 150 MeV/c similar to what was seen in ref \cite{12c_gamma_2N_2}.

\section{Conclusions}

This experiment has measured exclusive (e,e$'$pn)
cross sections for the first time. A large solid-angle array of time-of-flight scintillators was used to detect
neutrons and to measure their energies.
 The energy resolution was good enough to separate groups of
states in the residual $^{14}$N nucleus, but not individual states.
The two groups of states observed were found to have different missing momentum dependencies. Theoretical calculations of the missing momentum spectrum for the low energy group of states are unable to
reproduce the measured strength at low missing momenta.

\section*{Acknowledgements}

The authors would like to thank the staff of the Institut f\"{u}r Kernphysik
in Mainz for providing the facilities in which this experiment took
place. This work was sponsored by UK Engineering and Physical Sciences
Research Council and the Deutsche Forschungsgemeinschaft.

\bibliographystyle{unsrt}
\addcontentsline{toc}{section}{\refname}\bibliography{references}

\end{document}